\documentclass[12pt,a4paper,oneside]{article}
\usepackage{graphicx}
\usepackage{color}
\usepackage{booktabs}
\usepackage{epsfig}
\usepackage{multirow}
\usepackage[english]{babel}
\usepackage{amsmath}
\usepackage{amssymb}
\usepackage[latin1]{inputenc}
\usepackage{fancyhdr}
\usepackage{indentfirst}
\usepackage{newlfont}
\usepackage{latexsym}
\usepackage[latin1]{inputenc}
\usepackage{fancyhdr}
\usepackage{booktabs}
\usepackage[hang,bf,small]{caption}
\usepackage{microtype}
\usepackage[a4paper,top=4cm,left=4cm]{geometry}

\def\a{\alpha}
\def\b{\beta}

\def\g{\gamma}
\def\l{\lambda}

\def\e{\epsilon}

\newtheorem{teo}{Theorem}[section]
\newtheorem{lem}{Lemma}[section]

\newtheorem{nas}{Corollary}[section]
\newtheorem{defini}{Definition}[section]

\newcommand{\LeftEqNo}{\let\veqno\leqno}

\title{The Clebsch System}
\author{F. Magri$^{1}$, T. Skrypnyk$^{1,2}$ \\
\normalsize{$^{1}$Dipartimento di Matematica e Applicazioni- Università di Milano Bicocca}\\
\normalsize{$^{2}$Bogolyubov Institute for Theoretical Physics, Kiev, Ukraine}\\
\normalsize{franco.magri$@$unimib.it, taras.skrypnyk@unimib.it}\\
}
\date{}
\begin{document}
\maketitle
\begin{abstract}
The Clebsch system is one of the few classical examples of rigid bodies whose equations of motion are known to be integrable in the sense of Liouville. The explicit solution of its equations of motion, however,  is particularly hard, and it has defeated many attempts in the past. In this paper we present a novel and rather detailed study of these equations  of motion. Our approach is based on an improved version of the method originally used, in 1889, by Sophia Kowalewski to solve the equations of motion of the top bearing her name. We improve her method in two important points, and we clarify that it concerns a class of dynamical systems which is wider than the class of Hamiltonian systems which are integrable in the sense of Liouville. We use the improved version of the method by Kowalewski to prove two results. First, without using the Hamiltonian structure of the Clebsch system, we show that the integration of the equations of motion may be achieved by computing four Abelian integrals. Next, taking into
account its Hamiltonian structure, we show that two quadratures are sufficient to compute a complete integral of its Hamilton-Jacobi equation. In this way, the process of solution of the equations of motion of  the Clebsch system is clarified both from the standpoint of Abel and from the standpoint of Jacobi.
\end{abstract}

\section{Introduction}

The Clebsch system is one of the few cases of motion of a rigid body which are known to be integrable in the sense of Liouville. The other cases are the Euler, Lagrange, and Kowalewski tops and the Steklov system.

The Clebsch system is rather old, having being discovered by Clebsch in 1871  \cite{Clebsch}. During its long history, the way of looking at this system has progressively shifted in time. If at the beginning it was nothing more than a particular form of the Euler-Poisson equations of a rigid body, in recent times it has become customary to define the Clebsch system as a 5-parametric family of Hamiltonian vector fields on the dual of the Lie algebra of the group of motions of the Euclidean space $\mathbb{E}_{3}$. These vector fields are defined by six ordinary differential equations, usually written in the vector form
\begin{align*}
\begin{split}
\dot{\vec{S}}&= \frac{\partial H}{\partial \vec{S}} \times\vec{S}+   \frac{\partial H}{\partial \vec{T}}\times \vec{T}\\
\dot{\vec{T}}&= \frac{\partial H}{\partial \vec{S}} \times \vec{T}
,
\end{split}
\end{align*}
with the understanding that the Hamiltonian function
\begin{equation*}
H= \sum_{\a=1}^3m_{\a}S^2_{\a}+\sum_{\a=1}^3 n_{\a}T^2_{\a}
\end{equation*}
is a quadratic function of the components of the vectors $\vec S $ and $\vec T $ , with coefficients that satisfy the constraint
\begin{equation*}
\frac{n_{1}-n_{2}}{m_{3}} + \frac{n_{2}-n_{3}}{m_{1}} +\frac{n_{3}-n_{1}}{m_{2}} = 0 .
\end{equation*}

The way of looking at the Clebsch system adopted in this paper is an intermediate form between these two viewpoints. In this paper the Clebsch system is regarded as a pair of commuting vector fields $X$ and $Y$ on $\mathbb{R}^{6}$, depending on three arbitrary parameters $ j_{1}, j_{2}, j_{3}$ , and possessing four common integrals of motion $ I_{1},I_{2}, I_{3}, I_{4}$ . The vector fields are defined by the differential equations

\begin{equation}
\label{ClebEq}
\begin{aligned}
\dot{S}_{1}&=(j_{3}-j_{2})T_{2}T_{3}, \qquad &\dot{S}_{2}&=(j_{1}-j_{3})T_{3}T_{1}, \qquad &\dot{S}_{3}&=(j_{2}-j_{1})T_{1}T_{2},\\
\dot{T}_{1}&=S_{2}T_{3}-S_{3}T_{2}, \qquad &\dot{T}_{2}&=S_{3}T_{1}-S_{1}T_{3}, \qquad &\dot{T}_{3}&= S_{1}T_{2}-S_{2}T_{1},
\end{aligned}
\end{equation}

 for $X$,  and

\begin{align}
\label{ClebEq'}
\begin{split}
{S}'_{1}&= j_{1}(j_{3}-j_{2})T_{2}T_{3}+
(j_{2}-j_{3})S_{2}S_{3},\\
{S}'_{2}&= j_{2}(j_{1}-j_{3})T_{3}T_{1}+
(j_{3}-j_{1})S_{3}S_{1},\\
{S}'_{3}&= j_{3}(j_{2}-j_{1})T_{1}T_{2}+
(j_{1}-j_{2})S_{1}S_{2},\\
{T}'_{1}&= j_{2}S_{2}T_{3}-j_{3}S_{3}T_{2},\\
{T}'_{2}&= j_{3}S_{3}T_{1}-j_{1}S_{1}T_{3},\\
{T}'_{3}&= j_{1}S_{1}T_{2}-j_{2}S_{2}T_{2},
\end{split}
\end{align}

for  $Y$. Their common  integrals of motion are

\begin{equation}
\begin{aligned}
 I_1&=\sum\limits_{\a=1}^3T_{\a}S_{\a},\qquad
 &I_2&=\sum\limits_{\a=1}^3T^2_{\a},\\
I_3&=\sum_{\a=1}^3S^2_{\a}+\sum_{\a=1}^3((j_1+j_2+j_3)-j_{\a})T^2_{\a},\qquad
&I_4&=\sum_{\a=1}^3j_{\a}S^2_{\a}+\sum_{\a=1}^3\frac{j_1j_2j_3}{j_{\a}}T^2_{\a}.\\
\end{aligned}
\end{equation}
Nothing is lost by adopting this point of view. Indeed the 5-parametric family of Clebsch vector fields  is readily recovered from $X$ and $Y$ by taking their linear span according to the equation
\begin{equation*}
\text{Clebsch vector fields} =  \lambda X + \mu Y  .
\end{equation*}
The vector field so constructed corresponds to the Hamiltonian $ H $ whose coefficients are
\begin{align*}
\begin{split}
m_{\a}&= \lambda+\mu j_{\alpha},\\
n_{\a}&= \lambda(j_{\beta}+j_{\gamma})+\mu j_{\beta}j_{\gamma}    .
\end{split}
\end{align*}
\bigskip

The qualifying feature of the present point of view is that it encompasses all the relevant informations about the symmetries and the conservation laws of the Clebsch system, but not its Hamiltonian structure. The phase space $\mathbb{R}^{6}$ is not required to be endowed with any Poisson bracket, and therefore one cannot set any correspondence between symmetries and conservation laws. The noticeable property is that , as shown in the next section, this correspondence is not necessary for integrating the equation of motion of the Clebsch system, contrary to a diffuse belief. All that is required to perform the integration is the existence of a complementary number of symmetries and conservation laws, whatever their origin may be. This geometrical setting is so peculiar to deserve a name.
\begin{defini}
An integrable system in the sense of Lie is a collection of $k$ commuting vector fields $(X_{1}, X_{2}, \cdots  , X_{k} )$ on a manifold $M$ of dimension $n$, together with a collection of $(n - k) $ scalar functions $( I_{1}, I_{2}, \cdots , I_{n-k}) $ which are constants along the leaves of the foliation $\mathcal{F}$ spanned by the above vector fields. The manifold $M$ is not required to be a Poisson manifold.
\end{defini}

Two are the problems tackled in this paper. The first is to show a strategy to  bring to  $ " quadratures" $ any system which is integrable in the sense of Lie, at least in the specific case  $ ( n=6, k=2 ) $. The second is to work out in detail this strategy for the particular example of the Clebsch system.

In this introduction we present the ideas which are behind our strategy in a qualitative way. The main object to be considered is the foliation $\mathcal{F}$. It clearly reminds the Lagrangean foliation of an  Hamiltonian integrable system with two degrees of freedom, but we insist to stress that this Hamiltonian setting is only a rather specialized case of the more general setting envisaged in the approach of Lie. The four integrals of motion $ ( I_{1}, I_{2}, I_{3}, I_{4} )$ play the role of coordinates on the space of leaves. To introduce a pair of additional coordinates $x_{1}$ and $x_{2}$ on each leave of $\mathcal{F}$, we look at the zeroes of a second-order algebraic equation
\begin{equation}
\label{KowP}
Ex^2+Fx+G=0,
\end{equation}
whose coefficients are functions on $M$ to be properly selected. This equation will be referred to as the \emph{characteristic equation} attached to the foliation $\mathcal{F}$. Following Sophia Kowalewski , we focus our attention on a special class of characteristic equations, which are said to enjoy the Kowalewski's property.

\begin{defini}
 The characteristic equation of the foliation $\mathcal{F}$ enjoys the Kowalewski's property if there exists a pair of functions $\psi_1(x_1, I ), \psi_2(x_2,I )$ ,  that do not depend on  $x_{2}$ and $x_{1}$ respectively, such that the equations of motion of the vector fields $X$ and $Y$, in the coordinate system $( x_{1}, x_{2}; I_{1}, I_{2}, I_{3}, I_{4} )$ , take the form
\begin{align}
\label{preAbX}
\begin{split}
\frac{\dot{x}_1}{\psi_1(x_1,I)}+
\frac{\dot{x}_2}{\psi_2(x_2,I)}&=0,\\
\frac{x_1\dot{x}_1}{\psi_1(x_1,I)}+
\frac{x_2\dot{x}_2}{\psi_2(x_2,I)}&=1,
\end{split}
\end{align}

and

\begin{align}
\label{preAbY}
\begin{split}
\frac{x_1'}{\psi_1(x_1,I)}+
\frac{x_2'}{\psi_2(x_2,I)}&=1,\\
\frac{x_1 x_1'}{\psi_1(x_1,I)}+
\frac{x_2 x_2'}{\psi_2(x_2,I)}&=0,
\end{split}
\end{align}
\noindent
respectively. If the functions $\psi_1(x_1,I)$ and $\psi_2(x_2,I)$ are, as we admit,the restrictions to the algebraic curves
\begin{align}
\begin{split}
\g_1    &:            P_{1}( x_{1}, y_{1} )  =0\\
\g_2   &:            P_{2}( x_{2}, y_{2} )  =0
\end{split}
\end{align}
of two rational functions $ R_{1}(x_{1}, y_{1}) $ and $ R_{2}(x_{2}, y_{2}) $, the above requirement means that the equations of motion of the vector field $X$, for instance, can be written in the Abel form
\bigskip
\begin{align}
\label{Ab}
\begin{split}
\frac{\dot{x}_1}{R_{1}(x_{1}, y_{1}) }+
\frac{\dot{x}_2}{R_{2}(x_{2}, y_{2}) }&=0,\\
\frac{x_1\dot{x}_1}{R_{1}(x_{1}, y_{1}) }+
\frac{x_2\dot{x}_2}{R_{2}(x_{2}, y_{2}) }&=1.
\end{split}
\end{align}

\bigskip
A similar result holds for $Y$. Hence the solutions of the equations of motion of both $X$ and $Y$ may be obtained by evaluating four Abelian integrals. This is the sense of the Kowalewski's property.
\end{defini}
\begin{flushright}
$\square$
\end{flushright}

For brevity, the left-hand side of a characteristic equation enjoying the Kowalewski's property will be referred to as a \emph{K-polynomial}. In Sec. 2 we identify a special class of K-polynomials through a couple of differential equations which are satisfied by their coefficients $ (E, F, G ) $. These conditions concern the derivatives of the functions $ (E, F, G ) $ along the vector fields $( X, Y ) $. They are a particular instance of two conditions that hold for any dynamical system which is integrable in the sense of Lie. Once these conditions are satisfied, one is sure that the equations of motion of any vector field which is tangent to the foliation $\mathcal{F}$ and commute with $X$ and $Y$  can be put in the Abel form. These conditions are  the first piece of information that we add to the method used by Kowalewski in 1889. When applied to the Clebsch system, they allow to discover the following result.

\begin{teo}
Let $ v( S_{\a}, T_{\a}) $ be the function on the phase space of the Clebsch system implicitly defined by the equation
\begin{equation}
\label{Vin}
\sum\limits_{\a=1}^3 c_{\a} (v_{\a}S_{\a}+v_{\b}v_{\g}T_{\a})=0,
\end{equation}
where the symbols $v_{\a}$ and  $c_{\a}$ are shorthand notations for the square roots  of  $v+j_{\a} $ and of  $j_{\b}-j_{\g}$ respectively . Hence
\begin{align}
\begin{split}
 v^2_{\a}&=v+j_{\a},\\
c^2_{\a}&=j_{\b}-j_{\g} ,
\end{split}
\end{align}
where $(\a, \b, \g)$ is an even permutation of $(1, 2, 3).$ Use the function $v$ to construct the auxiliary functions
\begin{align}
\label{efg}
\begin{split}
E&=\sum\limits_{\a=1}^3c_{\a}v_{\a}T_{\a},\\
F&=\sum\limits_{\a=1}^3c_{\a}(v_{\b}v_{\g}S_{\a}+(j_{\b}+j_{\g})v_{\a}T_{\a}),\\
G&=\sum\limits_{\a=1}^3c_{\a}(v_{\b}v_{\g}j_{\a}S_{\a}+j_{\b}j_{\g}v_{\a}T_{\a}),
\end{split}
\end{align}
and with them the characteristic equation
\begin{equation*}
\label{KowP}
Ex^2+Fx+G=0.
\end{equation*}
Attach this equation to the foliation defined by the vector fields $X$ and $Y$ of  Clebsch.Then the following claims hold true:
\begin{enumerate}
\item The above characteristic equation  enjoys the Kowalewski's property relative to the vector fields $X$ and $Y$ of  Clebsch .
\item  The equations of motion of the whole 5-parametric family of vector fields of Clebsch take, simultaneously, the Abel form in the coordinates $x_1$ and $x_2$ .
\item The Abel equations associated with distinct vector fields of the family contain the same rational functions, and differ only by the values of the constants appearing on the right-hand side.
\item  The rational functions of the Clebsch family are
\begin{align}
\begin{split}
R_{1}( x_{1}, y_{1} ) &= - 8y_1(x_1+j_1)(x_1+j_2)(x_1+j_3)\\
R_{2}( x_{2}, y_{2} ) &=  2 \dfrac{I_1^2}{y_2^3} - 2y_2(x_2+j_1)(x_2+j_2)(x_2+j_3).
\end{split}
\end{align}
\item  These rational functions must be restricted over the same algebraic curves $\g_{1}$ and $\g_{2}$ .
\item The algebraic curves of the Clebsch family have the following equations
\end{enumerate}
\begin{small}
\begin{align*}
4y_1^2(x_1+j_1)(x_1+j_2)(x_1+j_3)+(x_1^2I_2+x_1I_3+I_4)+2I_1\sqrt{(x_1+j_1)(x_1+j_2)(x_1+j_3)}&=0,\\
y_2^4(x_2+j_1)(x_2+j_2)(x_2+j_3)+ y_2^2(x_2^2I_2+x_2I_3+I_4)+I_1^2&=0.
\end{align*}
\end{small}

These informations specify the Abelian integrals which one has to compute, in principle, to work out the solutions of the equations of motion of the Clebsch system.
\begin{flushright}
$\square$
\end{flushright}
\end{teo}

The above results do not depend on the Hamiltonian structure of the
Clebsch system. Nevertheless, once this structure is taken into
account, they lead readily to construct  a complete integral of
the Hamilton-Jacobi equation associated  with the Clebsch system. Indeed, one may  prove the following additional result.

\begin{teo}
Let $W_1(x_1; I)$  and $W_2(x_2; I)$ be the general solutions of the ordinary differential equations
\begin{small}
\begin{align*}
\begin{split}
\frac{dW_1}{dx_1}&=\sqrt{-\Bigl(\frac{(x_1^2I_2+x_1I_3+I_4+2\sqrt{x_1+j_1}\sqrt{x_1+j_2}\sqrt{x_1+j_3}I_1)}{4
(x_1+j_1)(x_1+j_2)(x_1+j_3)}\Bigr)},\\
\frac{dW_2}{dx_2}&= \sqrt{\Bigl(\frac{
-(x_2^2I_2+x_2I_3+I_4)+\sqrt{(x_2^2I_2+x_2I_3+I_4)^2-4I_1^2(x_2+j_1)(x_2+j_2)(x_2+j_3)}}{2(x_2+j_1)(x_2+j_2)(x_2+j_3)}\Bigr)
} .
\end{split}
\end{align*}
\end{small}
Then the function $W(x_1,x_2; I)=W_1(x_1; I)+ W_2(x_2; I)$ is a complete integral of the Hamilton-Jacobi equations associated with all the vector fields of the Clebsch family.
\end{teo}
\begin{flushright}
$\square$
\end{flushright}

This result stems from the clarification of the relation between the Abel's and the Jacobi's forms of the equations of motion worked out in Sec. 2. This clarification is the second improvement brought to the method of Kowalewski.

The plan of the paper is rather simple. It  consists of four sections. Each of them is devoted to one step of the procedure outlined before . In Sec. 2 we present the improved version of the method by Kowalewski. In particular we present the conditions on the coefficients $(E, F, G ) $ of the characteristic equation which were missing in the approach of Kowalewski, and that define the class of K-polynomials which are of interest for the present paper.We also clarify the relationship between the Abel's and the Jacobi's forms of the equations of motion. In Sec. 3 we explain the origin of the characteristic equation of the Clebsch system. Finally, in Sec. 4, we work out the Abel's  and the Jacobi's form of the equations of motion of Clebsch, proving the Theorems 1.1 and 1.2. Finally we express the original mechanical variables $S_{\a}$ and $T_{\a}$ in terms of the coordinates $x_{1}$ and $x_{2}$ , of their conjugate momenta $p_{1}$ and $p_{2}$ , and of the Casimir functions $I_{1}$ and $I_{2}$ .  The paper ends with a short comparison with previous attempts of solving the equations of motion of the Clebsch system.

\section{The Kowalevski 's method and some of its corollaries}

In this section we present three theorems which serve as  basis of our approach to the Clebsch system. They specify a class of integrable dynamical systems and provide an explicit algorithm for
solving the corresponding equations of motion. The theorems are borrowed from a
set of unpublished notes of the first author on the geometry of integrable systems.

The geometrical setting has already been outlined in the Introduction. On a manifold $M$  of dimension $6$, we consider  two commuting vector field $X$ and $Y$, and a collection $(I_1, I_2, I_3, I_4)$ of functions which are constant along the leaves of the foliation $\mathcal{F}$ spanned by the vector fields. To this foliation we attach the characteristic equation
\begin{equation*}
Ex^2+Fx+G=0.
\end{equation*}
We assume that the following conditions hold true on an open subset $U$ of $M$, formed by a collection of leaves of $\mathcal{F}$:
\begin{align*}
\begin{split}
E &\neq 0,\\
F^2-4EG&\neq 0,\\
EdF\wedge dG +FdG\wedge dE+GdE\wedge dF &\neq 0,
\end{split}
\end{align*}
We also assume that
\begin{equation*}
\dot{x}_1 x_2'-x_1'\dot{x}_2 \neq 0 .
\end{equation*}
In this way one is guaranteed that the roots $x_1$, $x_2$ of the characteristic equation may be used as coordinates on the leaves of $\mathcal{F}$ sitting on $U$. These minor requirements, introduced just to avoid patologies, are subsequently completed by two major conditions, referred to as the \emph{Kowalewski's condition} $\textit{K1}$ and $\textit{K2}$ . Their role is explained by the following statement.

\begin{teo}
If the coefficients $E$, $F$, $G$ of the characteristic equation verify the pair of differential equations:
\begin{equation}\LeftEqNo
\tag{\textit{K1}}
EF'-FE'-(E\dot{G}-G\dot{E})=0,
\end{equation}
\begin{equation}\LeftEqNo
\tag{\textit{K2}} EG'-GE'+(G\dot{F}-F\dot{G})=0,
\end{equation}
the characteristic equation enjoys the Kowalevski' s property. Consequently, its roots $x_1$ and
$x_2$ allow  to put the equations of motions of the vector fields $X$ and $Y$ in the Abel form .
\end{teo}

{\textit Proof.}  We divide the proof in three parts. First of all, we notice that the assumptions $\textit{K1}$ and  $\textit{K2}$ imply that the roots $x_1$ and $x_2$ satisfy the differential equations:
\begin{equation}\LeftEqNo
\tag{\textit{K1'}}
x_{1}'+x_2\dot{x_1}=0,
 \end{equation}
\begin{equation}\LeftEqNo
\tag{\textit{K2'}}
x_2'+x_1\dot{x_2}=0.
 \end{equation}
This claim easily follows from the identities:
\begin{align*}
x_1+x_2&=-\frac{F}{E},\\
x_1x_2&=\frac{G}{E}
\end{align*}
and their differential consequences:
\bigskip
\begin{align*}
(x_1'+x_2\dot{x}_1)+(x_2'+x_1\dot{x}_2)&=\left(\frac{G}{E}\right)^{\bullet}-\left(\frac{F}{E}\right)',\\
x_2(x_1'+x_2\dot{x}_1)+x_1(x_2'+x_1\dot{x}_2)&=\left(\frac{G}{E}\right)'-
\frac{F}{E}\left(\frac{G}{E}\right)^{\bullet}+ \frac{G}{E}\left(\frac{F}{E}\right)^{\bullet}.
\end{align*}
\bigskip

The new  form of the Kowalewski's conditions is used so often henceforth that it is suitable to give it a name. The two differential equations satisfied by the roots of the characteristic equation will be simply referred to as the conditions $\textit{K1'}$ and $\textit{K2'}$, and called the second form of the Kowalewski's conditions.

As a second step, we introduce the functions
\begin{align}
\begin{split}
\psi_1 :=&(x_1-x_2)\dot{x}_1,\\
\psi_2 :=&(x_2-x_1)\dot{x}_2.
\end{split}
\end{align}
They are used to compute the components of  the vector fields $\frac{\partial }{\partial x_1}$ and $\frac{\partial }{\partial x_2}$ , of the canonical basis associated with  the coordinates $x_1$ and $x_2$, along the vector fields $X$ and $Y$. On account of the second form of the Kowalewski's condition, one readily obtains the formulas
\begin{align*}
\begin{split}
\psi_1 \frac{\partial}{\partial x_1}&=Y+x_1 X,\\
\psi_2 \frac{\partial}{\partial x_2}&=Y+x_2 X.
\end{split}
\end{align*}

The third and final step is to notice that the vector field $Y+x_1 X$ and
$Y+x_2 X$ commute. Indeed, one finds:
\begin{equation*}
[Y+x_1X,Y+x_2X]=((x_2'+x_1\dot{x}_2)- (x_1'+x_2\dot{x}_1))X,
\end{equation*}
and one concludes that the commutator vanishes owing to the identities $\textit{K1'}$ and $\textit{K2'}$.
This commutation relation entails
\bigskip
\begin{equation*}
\left[\psi_1 \dfrac{\partial}{ \partial x_1},\psi_2 \dfrac{\partial}{\partial x_2}\right]  = 0 .
\end{equation*}
But this is possible if and only if
\begin{equation*}
\frac{\partial \psi_2}{\partial x_1}=0, \  \frac{\partial \psi_1}{ \partial x_2}=0.
\end{equation*}
The conclusion is that the function $\psi_1$ does not depend on $x_{2}$, and that the  function $\psi_2$ does not depend on $x_{1}$, as required by the Kowalewski's property. It remains thus proved that the Kowalewski's condition  $\textit{K1}$ and  $\textit{K2}$ entail that the characteristic equation enjoys the Kowalewski's property.
\begin{flushright}
$\square$
\end{flushright}

It is instructive to see immediately  the conditions  $\textit{K1}$ and $\textit{K2}$, or equivalently the conditions $\textit{K1'}$ and $\textit{K2'}$, at work in a classical example.

\bigskip
\noindent
\textbf{Example}. The Kowalewski top perfectly fits the geometric scheme considered in this paper. Indeed , it can be dealt with as a dynamical system on $\mathbb{R}^{6}$  which is integrable in the sense of Lie. In other words, one may interpret the famous paper \textit{" Sur le mouvement d' un corps rigide autour d' un point fixe"} \cite{Kowal} as the study of a very special two-dimensional foliation on $\mathbb{R}^{6}$, spanned by two commuting vector fields $X$ and $Y$ and possessing four integrals $I_1$, $I_2$, $I_3$, $I_4$, without specifying any Poisson bracket on $\mathbb{R}^{6}$.  This point of view is actually strictly adherent to the work of Kowalewski , who could not be aware of the Hamiltonian structure of her top, and that consequently never mention nor use this concept in her paper.  In discussing this example, our aim is to show that the conditions $\textit{K1'}$ and $\textit{K2'}$ are of help in understanding  the deep reasons of the success of the computations performed by Kowalewski.

The vector fields $X$ and $Y$ of Kowalewski are defined by two sets of differential equations. The differential equations of the vector field $X$ which describes the motion of the top are:

\begin{equation*}
\begin{aligned}
 \dot{L_1}&=\frac{1}{2}L_2L_3,& \dot{L_2}&=-\frac{1}{2}L_1L_3-y_3,&
 \dot{L_3}&=y_2,\\
 \dot{y_1}&=L_2y_2-\frac{1}{2}L_2y_3,&
 \dot{y_2}&=\frac{1}{2}L_1y_3-L_3y_1,&
 \dot{y_3}&=\frac{1}{2}L_2y_1-\frac{1}{2}L_1y_2.
\end{aligned}
\end{equation*}
The differential equations of the symmetry $Y$ are:
\begin{align*}
\begin{split}
L_1'&=\frac{1}{2}k_1L_2L_3-k_2(\frac{1}{2}L_1L_3+y_3),\\
L_2'&=\frac{1}{2}k_1(\frac{1}{2}L_1L_3+y_3)+\frac{1}{2}k_2L_2L_3,\\
L_3'&=-k_1L_1L_2+\frac{1}{2}k_2(L_1^2-L_2^2)+(k_2y_1-k_1y_2),\\
y_1'&=\frac{1}{2}(k_1L_2-k_2L_1)y_3,\\
y_2'&=\frac{1}{2}(k_1L_1+k_2L_2)y_3,\\
y_3'&=-\frac{1}{2}k_1(L_1y_2+L_2y_1)+
\frac{1}{2}k_2(L_1y_1-L_2y_2).
\end{split}
\end{align*}
where  $k_1$ and $k_2$ are shorthand notations for the functions:
\begin{equation*}
k_1=(L_1^2-L_2^2)+4y_1,\ k_2=2L_1L_2+4y_2.
\end{equation*}
As  integrals of the foliation  Kowalewski has chosen  the functions :
\begin{align*}
I_1&=L_1y_1+L_2y_2+L_3y_3,\\
I_2&=y_1^2+y_2^2+y_3^2=1,\\
I_3&=\frac{1}{4}(L_1^2 +L_2^2+2L_3^2)-y_1,\\
I_4&=\frac{1}{8} (k_1^2+ k_2^2),
\end{align*}

This choice is the most natural from the viewpoint of Hamiltonian mechanics. Nevertheless, we emphasize again that it is quite useful to keep the freedom to choose the vector fields and the integrals of the foliation independently, insofar the vector fields commute. This freedom can be exploited to simplify the process of integration of the equations of motion.

To solve the equations of motion of the top, Kowalevski attached to the foliation spanned by the vector fields $X$ and $Y$ the algebraic equation
\begin{equation*}
(z_1-z_2)^2 x^2- 2R(z_1,z_2)x-R_1(z_1,z_2)=0,
\end{equation*}
called the fundamental equation of Kowalevski by Golubev \cite{Golubev}. In this equation the variables are
\begin{align*}
z_1&=\frac{1}{2}(L_1+iL_2)\\
z_2&=\frac{1}{2}(L_1-iL_2)
\end{align*}
and the coefficients are the polynomials :
\bigskip
\begin{small}
\begin{align*}
R(z_1,z_2)&=-z_1^2z_2^2+2I_1z_1z_2+I_3(z_1+z_2)+I_4-\frac{1}{2}I_2,\\
R_1(z_1,z_2)&=-2I_1z_1^2z_2^2-(I_4-\frac{1}{2}I_2)(z_1+z_2)^2-2I_3(z_1+z_2)z_1z_2+2I_1(I_4-\frac{1}{2}I_2)-I_3^2.
\end{align*}
\end{small}

Then she proved, by a direct inspection of the behavior of the roots $x_1$ and $x_2$ along the trajectories of the vector field $X$,  that the fundamental equation enjoys the Kowalewski's property.This result has ever been considered as rather mysterious, since nothing in the computations performed by Kowalewski could suggest such a kind of outcome. In other words, the Abel form of the equations of motion comes out  as an unexpected result. In the light of the previous Theorem, we are in a position to justify \emph{a priori} the result by Kowalewski, without ever using or computing the coordinates $x_1$ and $x_2$ in the style of Kowalewski.

To this end, we use the geometric interpretation of the fundamental equation of Kowalewski presented in \cite{Magri} . In this paper, the fundamental equation is interpreted as the characteristic equation of a tensor field $\mathcal{K}$ of  type $(1,1)$ . This tensor field is defined on the leaves of the foliation $\mathcal{F}$ by its action on the canonical basis $dz_1$ and $dz_2$ associated with the mechanical coordinates $z_1$ and $z_2$. To wit , the definition of $\mathcal{K}$ is:
\bigskip
\begin{align*}
(z_1-z_2)^2\mathcal{K}dz_1&=R(z_1,z_2)dz_1+R(z_1,z_1)dz_2, \\
(z_1-z_2)^2\mathcal{K}dz_2&=R(z_2,z_2)dz_1+R(z_1,z_2)dz_2.
\end{align*}

This geometric standpoint allows to point out two hidden properties of the fundamental equations of Kowalewski in the form of  properties enjoyed by the related tensor field $\mathcal{K}$. As noticed in \cite{Magri}, this tensor field obeys two conditions. The first states that the spectral invariants of $\mathcal{K}$ satisfy the constraint:
\begin{equation*}
\mathcal{K} d(\mathrm{Tr}\mathcal{K}) =d(\mathrm{det}\mathcal{K}),
\end{equation*}
The second states that $\mathcal{K}$ maps $X$ into $Y$ according to the equation:
\begin{equation*}
Y+\mathcal{K}X=0.
\end{equation*}
(To avoid the annoying presence of a numerical coefficient in this equation, one has to suitably rescale the vector field $Y$ in this equation, by using the freedom of defining the vector fields $X$ and $Y$ independently of the integrals of motion mentioned above.) Together these two conditions mean that the eigenvalues of $\mathcal{K}$ satisfy the conditions  $\textit{K1'}$ and $\textit{K2'}$, a fact that can be easily checked without computing the eigenvalues of $\mathcal{K}$. Thus, according to the theorem proved before,  one can foresee that the equations of motion of the Kowalewski's top  can be brought into the Abel form, without computing explicitly this form. In this sense,  the theorem proved in this section gives an  intrinsic and \emph{a priori} explanation of the success of the attempt done by Kowalewski.
\begin{flushright}
$\square$
\end{flushright}

To complete the analysis of the integrability of the foliation $\mathcal{F}$, we must
  consider specifically the Hamiltonian case. The geometric setting is accordingly modified as follows.
  The 6-dimensional manifold $M$ is now endowed with a degenerate Poisson bracket, possessing two Casimir
  functions $C_1$ and $C_2$. Moreover, a pair of functions $H$ and $K$ is given on $M$  which are in involution
   with respect to the Poisson bracket. They define the pair of commuting Hamiltonian vector fields $X_H$ and $X_K$, which span a 2-dimensional Lagrangean foliation. The integrals of this foliation are the functions $I_1=C_1$, $I_2=C_2$, $I_3=H$, $I_4=K$. A characteristic equation is attached, as usual, to this foliation. The choice of the characteristic equation is restricted by  asking that its roots be in involution with respect to the Poisson bracket defined on $M$. This requirement demands that the coefficients $E$, $F$, $G$ satisfy  the additional constraint
\begin{equation}\LeftEqNo
\tag{\textit{K3}}
E\{F,G\}+G\{E,F\}+F\{G,E\}=0,
\end{equation}
which is equivalent to the condition
 \begin{equation}\LeftEqNo
 \tag{\textit{K3'}}
\{x_1,x_2\}=0
\end{equation}
on the roots.  This last condition  will be referred to as the second form of the third Kowalewski's condition , and called the condition \textit{K3'}. This geometric setting defines the following class of dynamical systems.

\begin{defini}. Let $\mathcal{F}$ be the 2-dimensional Lagrangean foliation spanned by the vector fields $X_H$ and $X_K$ on a Poisson manifold of dimension 6, endowed with a degenerate Poisson bracket possessing two Casimir functions  $C_1$ and $C_2$. If it is possible to attach to the foliation $\mathcal{F}$ a characteristic equation whose coefficients satisty the three Kowalewski's conditions $\textit{K1}$,$\textit{K2}$,$\textit{K3}$ the vector fields $X_H$ and $X_K$ are said to define a KCI system, that is an Hamiltonian system which is completely integrable in the sense of Kowalewski.
\end{defini}

A KCI system is automatically a system which is integrable in the sense of Lie, and therefore its equations of motion can be reduced to the Abel form. But this Abel form is rather special, as shown by the following remark.

\begin{lem}  The functions $\psi_1(x_1,H,K,C_1,C_2)$ and $\psi_2(x_2,H,K,C_1,C_2)$ , entering
into the Abel form of the equations of motion, contain the Hamiltonian functions $H$ and $K$ only in the form $K+xH$, that is
\begin{align*}
\psi_1(x_1,H,K,C_1,C_2)&\equiv \psi_1(x_1,Hx_1+K,C_1,C_2),\\
\psi_2(x_2,H,K,C_1,C_2)&\equiv \psi_2(x_2,Hx_2+K,C_1,C_2).
\end{align*}
\end{lem}
\textit{Proof.} By deriving condition $\textit{K3'}$ along the vector
fields $X_H$ and $X_K$, one obtains:
\begin{align*}
\{\dot{x}_1,x_2\}+\{{x}_1,\dot{x}_2\}&=0,\\
\{{x}_1',x_2\}+\{{x}_1,{x}_2'\}&=0.
\end{align*}
Due to conditions $\textit{K1'}$ and $\textit{K2'}$, these identities may also be written in the form:
 \begin{align*}
\{\dot{x}_1,x_2\}+\{{x}_1,\dot{x}_2\}&=0,\\
x_2 \{\dot{x}_1,x_2\}+x_1\{{x}_1,\dot{x}_2\}&=0.
\end{align*}
Since the roots $x_1$ and $x_2$ are distinct, the unique possibility is that:
\begin{equation*}
\{\dot{x}_1,x_2\}=0, \{{x}_1,\dot{x}_2\}=0.
\end{equation*}
Using the definition (13) of the functions $\psi_{1}, \psi_{2}$ we
obtain:
\begin{equation*}
\{\psi_1,x_2\}=0, \{\psi_2,x_1\}=0.
\end{equation*}
Let us evaluate the first Poisson bracket by using the Leibniz rule. We get:
\begin{equation*}
0=\{x_2,\psi_1\}=\frac{\partial \psi_1}{\partial H}\dot{x}_2+
\frac{\partial \psi_1}{\partial K}{x}_2'= (\frac{\partial
\psi_1}{\partial H}-x_1 \frac{\partial \psi_1}{\partial
K})\dot{x}_2.
\end{equation*}
In the open subset $U$, where the roots  $x_1$ and $x_2$ act as coordinates on the leaves of $\mathcal{F}$, there are not singular points of the vector field $X_H$. Therefore the above  identity implies
\begin{equation*}
\frac{\partial \psi_1}{\partial H}-x_1 \frac{\partial
\psi_1}{\partial K}=0.
\end{equation*}
Similarly one proves that
\begin{equation*}
\frac{\partial \psi_2}{\partial H}-x_2 \frac{\partial
\psi_2}{\partial K}=0.
\end{equation*}
These two partial differential equations prove the claim made in the Lemma.
\begin{flushright}
$\square$
\end{flushright}

The Lemma just proved is the key to solve the Hamilton equations of motion in the style of Jacobi. Indeed it provides a rather effective way of computing the momenta $p_1$, $p_2$ conjugated to the coordinates $x_1$ and $x_2$.

\begin{teo}
The functions
\begin{equation*}
 p_i=\Psi_{i}(x_i,x_iH+K,C_1,C_2)=\int\limits^{x_iH+K}_0
\frac{d\lambda}{\psi_i(x_i,\lambda,C_1,C_2)},
\end{equation*}
are the momenta  canonically conjugated to the roots
$x_1$ and $x_2$ of the characteristic equation.
\end{teo}
{\textit Proof.}  Let us first compute the Poisson brackets of $p_1$ with the coordinates $x_1$ and $x_2$. We notice that:
\bigskip
\begin{align*}
\{x_1,p_1\}&=\frac{\partial \Psi_1(x_1,X_1,C_1,C_2) }{\partial
x_1}\{x_1,x_1\}+ \frac{\partial \Psi_1(x_1,X_1,C_1,C_2) }{\partial
X_1}\{x_1,x_1H+K\}\\&
=\frac{\partial\Psi_1(x_1,X_1,C_1,C_2)
}{\partial X_1}(x_1'+x_1\dot{x}_1)\\
&=\frac{\partial
\Psi_1(x_1,X_1,C_1,C_2) }{\partial X_1}(x_1-x_2)\dot{x}_1\\
&=\frac{\partial\Psi_1(x_1,X_1,C_1,C_2) }{\partial X_1}\psi_1\\
&=1,
\end{align*}
owing to condition $\textit{K1'}$. Furthermore
\bigskip
\begin{align*}
\{x_2,p_1\}&=\frac{\partial \Psi_1(x_1,X_1,C_1,C_2) }{\partial
x_1}\{x_2,x_1\}+ \frac{\partial \Psi_1(x_1,X_1,C_1,C_2) }{\partial
X_1}\{x_2,x_1H+K\}\\
&=\frac{\partial\Psi_1(x_1,X_1,C_1,C_2)
}{\partial X_1}(x_2'+x_1\dot{x}_2)\\
&=0,
\end{align*}
owing to condition $\textit{K2'}$. In the analogous way one obtains  $\{x_1,p_2\}=0$,
$\{x_2,p_2\}=1$. Finally, let us calculate $\{p_1,p_2\}$. We have:
\bigskip
\begin{align*}
\{p_1,p_2\}&=\frac{\partial \Psi_2(x_2,X_2,C_1,C_2) }{\partial
x_2}\{p_1,x_2\}+ \frac{\partial \Psi_2(x_2,X_2,C_1,C_2) }{\partial
X_2}\{p_1,x_2H+K\}\\
&=-\frac{\partial\Psi_2(x_2,X_2,C_1,C_2)
}{\partial X_2}\{x_2H+K,
\Psi_1(x_1,X_1,C_1,C_2)\}\\
&=-\frac{\partial\Psi_2(x_2,X_2,C_1,C_2)
}{\partial X_2}\frac{\partial\Psi_1(x_1,X_1,C_1,C_2) }{\partial
x_1}\{x_2H+K, x_1\}+ \\&-\frac{\partial\Psi_2(x_2,X_2,C_1,C_2)
}{\partial X_2}\frac{\partial\Psi_1(x_1,X_1,C_1,C_2) }{\partial
X_1}\{x_2H+K, x_1H+K\}\\
&=\frac{\partial\Psi_2(x_2,X_2,C_1,C_2)
}{\partial X_2}\frac{\partial\Psi_1(x_1,X_1,C_1,C_2) }{\partial
x_1}(x_1'+x_2\dot{x}_1)+\\&+\frac{\partial\Psi_2(x_2,X_2,C_1,C_2)
}{\partial X_2}\frac{\partial\Psi_1(x_1,X_1,C_1,C_2) }{\partial
X_1}H(-x_2'+x_1'+x_2\dot{x}_1-x_1\dot{x}_2)\\
&=0.
\end{align*}
due to both conditions $\textit{K1'}$ and $\textit{K2'}$. This identity completes the proof of the Theorem.
\begin{flushright}
$\square$
\end{flushright}

It is worth to remark here a difference with the standard theory of separation of variables. In this last theory the task is to find simultaneously the coordinates and the momenta which allow to perform the separation of variables . Otherwise it becomes impossible, for instance, to use the Levi Civita separability conditions or to find the separation equations. On the contrary , in the approach based on the characteristic equation of Kowalewski, the problem is splitted in two parts. The true problem is to find the characteristic equation which enjoys the property of Kowalewski, and hence the coordinates $x_1$ and $x_2$. The momenta then follow. This splitting of the problem has great advantages both practical and theoretical. Once the momenta have been computed according to the procedure outlined in Theorem 2.2, the way for the solution of the Hamilton-Jacobi equations associated with the Hamiltonian vector fields  $X_H$ and $X_K$ is completely paved.

In  the following two sections of the paper our task will be to show how this integration scheme can be effectively implemented in the case of the Clebsch system.

\section{The Clebsch system is a KCI system}

This section has two purposes. The first is to prove that the functions $E$, $F$, $G$ attached to the Clebsch system in Sec. 1 satisfy the  Kowalewski's conditions $\textit{K1}$, $\textit{K2}$, and $\textit{K3}$. The second is to explain where these functions come from.

The task of checking the Kowalewski's conditions is rather trivial if one knows the functions $E$, $F$,$G$ and the vector fields $X$ and $Y$ .  The conditions are tensorial. Therefore the problem is to compute the derivatives of the functions along the vector fields in any preferred coordinate system, and to see if the Kowalewski's conditions are satisfied or not. Nevertheless, in the present case there is a computational difficulty  due to the fact that the functions $E$, $F$, $G$  are only implicitly defined. Let us analyse this difficulty. As a first step, one has to compute the derivatives of the function $v$ along $X$ and $Y$. By the Implicit Function Theorem
\begin{equation*}
\dot{v}=-2v_1v_2v_3\frac{\sum\limits_{\a=1}^3
c_{\a}({v}_{\b}v_{\g}\dot{T}_{\a}+v_{\a}\dot{S}_{\a})}{2Ev+F},
\end{equation*}
and therefore the best one can do it is to write $\dot{v}$ as a rational function of the
 variables ${S}_{\a}$,${T}_{\a}$, ${v}_{\a}$, $v$. The same is true for $v'$, and \emph{a fortiori} for
 $\dot{E}$, $\dot{F}$,  $\dot{G}$, ${E}'$, ${F}'$, ${G}'$. Consequently the left-hand sides of  the Kowalewski's
 conditions  are given by rational functions of the same variables. These variables, however, are not independent,
  but are related by four constraints. Therefore to check the Kowalewski's conditions it is necessary to show that
   the numerator of these rational functions belong to the ideal generated by the four constraints. This is a cumbersome
    problem of
    commutative algebra.
For this reason we shall follow a different procedure.

The alternative is to use the second form of the Kowalewski's conditions. This approach  requires to overcome two difficulties. The first is to compute the roots  $x_1$ and $x_2$ . The second is to compute their derivatives along the vector field $X$ and $Y$. The solution of the first problem is easy. It is provided by the following Lemma.
\begin{lem}
The roots of the Clebsch's characteristic equation are $x_1=v$ and $x_2=-v-F/E$.
\end{lem}
{\textit Proof.}  From the definition of the functions $E$, $F$, $G$ the following identity readily follows:
\begin{equation*}
Ev^2+Fv+G=v_1v_2v_3\sum\limits_{\a=1}^3
c_{\a}(v_{\b}v_{\g}T_{\a}+v_{\a}S_{\a}).
\end{equation*}
The right-hand side is zero due to the constraint  on $v$. Hence
$x_1=v$ is a first solution of the Clebsch's characteristic equation. The second solution is obviously $x_2=-v-F/E$.
\begin{flushright}
$\square$
\end{flushright}

More elaborate is the solution of the second problem. There are several techniques that can be adopted to evaluate the derivates of the roots $x_1$ and $x_2$ along the vector fields $X$ and $Y$. The thecnique which we adopt here is to build, in three steps, an intermediate coordinate system containing the roots $x_1$ and $x_2$, in order to exploit then the usual transformation formulas of tensor calculus. (The present coordinates are intermediate between the original mechanical coordinates  $S_{\a}$, $T_{\a}$ and the final coordinates $(x_1, x_2, I_1, I_2, I_3, I_4 ) $ ). The main tool to construct the intermediate coordinates are the definitions
\begin{align*}
\begin{split}
E&=\sum\limits_{\a=1}^3c_{\a}v_{\a}T_{\a},\\
F&=\sum\limits_{\a=1}^3c_{\a}(v_{\b}v_{\g}S_{\a}+(j_{\b}+j_{\g})v_{\a}T_{\a}),\\
G&=\sum\limits_{\a=1}^3c_{\a}(v_{\b}v_{\g}j_{\a}S_{\a}+j_{\b}j_{\g}v_{\a}T_{\a}),
\end{split}
\end{align*}
of the coefficients of the characteristic equation attached to the Clebsch system. They are used to replace the coordinates $T_{\a}$ by the coordinates $E$, $F$, $G$. This step requires a few comments. The relations between the coordinates  $E$, $F$, $G$ and the coordinates $T_{\a}$ are rather complicated, because of the presence of the function $v$ within the coefficients of the transformation.
This function depends irrationally  on $S_{\a}$, $T_{\a}$, and therefore it may seem difficult to solve the above relations with respect to the variables  $T_{\a}$.  Nevertheless , one may notice that the previous Lemma tells us that  $v$ is a function of $E$, $F$, $G$. Therefore the same equations may also be treated as a linear system in $T_{\a}$, with coefficients that are complicated functions of $E$, $F$, $G$. This is the standpoint which we shall adopt henceforth.\\

The second step  is to replace the coordinates $F$, $G$ by the coordinates $x_1=v$ and $x_2=-v-F/E$. Again the previous Lemma allows us to write the transformation  in the more explicit form
\begin{align*}
\begin{split}
 E&=\sum\limits_{\a=1}^3c_{\a}\sqrt{(x_1+j_{\a})}T_{\a},\\
F&=-E(x_1+x_2)=\sum\limits_{\a=1}^3c_{\a}(\sqrt{(x_1+j_{\b})}\sqrt{(x_1+j_{\g})}S_{\a}+(j_{\b}+j_{\g})\sqrt{(x_1+j_{\a})}T_{\a}),\\
G&=Ex_1x_2=\sum\limits_{\a=1}^3c_{\a}(\sqrt{(x_1+j_{\b})}\sqrt{(x_1+j_{\a})}j_{\a}S_{\a}+j_{\b}j_{\g}\sqrt{(x_1+j_{\a})}T_{\a}).
\end{split}
\end{align*}
These formulas express the old coordinates $T_{\a}$ as explicit functions of the six coordinates $(S_{\a}, x_1, x_2, E )$. For further convenience, the last coordinate $E$ is henceforth replaced by $e=\frac{E}{c_1c_2c_3}$.
\bigskip

The third step is to eliminate the coordinates $S_{\a}$ as well.
Here we have much more freedom. The only guiding principle is that
the new coordinates must simplify as much as possible the form of
the equations of motion, and that the transformation of
coordinates could be inverted in a reasonable simple manner. We
have found that the variables
\begin{align*}
\begin{split}
l_1&=\frac{1}{c_1c_2c_3}  \sum\limits_{\a=1}^3
c_{\a}(x_1+j_{\a})\sqrt{(x_1+j_{\a})}S_{\a},\\
l_2&=\frac{1}{c_1c_2c_3}  \sum\limits_{\a=1}^3
c_{\a}(x_2+j_{\a})\sqrt{(x_1+j_{\a})}S_{\a},\\
l_3&=\frac{1}{c_1c_2c_3}  \sum\limits_{\a=1}^3
c_{\a}\sqrt{(x_1+j_{\b})}\sqrt{(x_1+j_{\g})}S_{\a}
\end{split}
\end{align*}
are particularly appropriate to the purpose.
\bigskip

The six equations just written define the intermediate coordinates $(x_1, x_2, e, l_1, l_2, l_3 )$ used in this section. The inverse transformation is
\bigskip
\begin{footnotesize}
\begin{align}\label{reconstr}
\begin{split}
S_{\a}=&{\frac{c_{\a}}{c_1c_2c_3}}(
\sqrt{x_1+j_{\a}}\frac{(j_{\b}+j_{\g}+x_1+x_2)}{x_1-x_2}l_1-
\sqrt{x_1+j_{\a}}\frac{(j_{\b}+j_{\g}+2x_1)}{x_1-x_2}l_2-
\sqrt{x_1+j_{\b}}\sqrt{x_1+j_{\g}}l_{3}),\\
\bigskip
T_{\a}=&{\frac{c_{\a}}{c_1c_2c_3}}(
\frac{\sqrt{x_1+j_{\b}}\sqrt{x_1+j_{\g}}}{x_1-x_2}(l_{1}-l_2)-
\sqrt{x_1+j_{\a}}l_3-\sqrt{x_1+j_{\a}}(x_2+j_{a})e).
\end{split}
\end{align}
\end{footnotesize}
\bigskip
Now it is an elementary problem of tensor calculus to evaluate the form of the equations of motion in the intermediate coordinates.  The result is presented in the following Lemma.

\begin{lem}
In the intermediate coordinates the equations of motion of the
vector fields $X$ are:
\begin{small}
\begin{align}
\begin{split}
(x_1-x_2)\dot{x}_1&=-4\bigl((x_1+j_1)(x_1+j_2)(x_1+j_3)
e+l_1\sqrt{(x_1+j_1)(x_1+j_2)(x_1+j_3)}\bigr),\\
(x_1-x_2)\dot{x}_2&= 2\bigl((x_2+j_1)(x_2+j_2)(x_2+j_3)e -
\frac{l_2^2}{e}),
\end{split}
\end{align}
\end{small}
and those of the vector field $Y$ are:
\begin{small}
\begin{align}
\begin{split}
(x_1-x_2){x}'_1&=4x_2\bigl((x_1+j_1)(x_1+j_2)(x_1+j_3)
e+l_1\sqrt{(x_1+j_1)(x_1+j_2)(x_1+j_3)}\bigr),\\
(x_1-x_2){x}'_2&= -2x_1 \bigl((x_2+j_1)(x_2+j_2)(x_2+j_3)e -
\frac{l_2^2}{e}\bigr).
\end{split}
\end{align}
\end{small}
\end{lem}

\textit{Proof.} The proof is obtained by implementing the standard transformation rule of tensor calculus.
\hfill{$\square$}
\bigskip

This Lemma is the gateway for the Abel form of the equations of motion of Clebsch. Even before computing the explicit form of the equations of motion in the final coordinates, we can be sure that they take the Abel form.

\begin{teo}
The equations of motion of the Clebsch system can be brought into the Abel form.
\end{teo}
\textit {Proof.} By the previous Lemma, the roots of the characteristic equation attached to the Clebsch system satisfy  the constraints
\begin{align*}
x_1'+x_2\dot{x_1}&=0\\
 x_2'+x_1\dot{x_2}&=0
 \end{align*}
The claim then follows from Theorem 2.1.
\begin{flushright}
$\square$
\end{flushright}

After the Abel form,  let us consider also the Jacobi form of  the equations of motion  of the Clebsch system. For that, we must  take into account the Hamiltonian structure of these equations, and therefore the Lie-Poisson bracket of $e^*(3)$:
\begin{align*}
\{ S_{\a}, S_{\b}\}&=\e_{\a\b\g}S_{\g}, \\
\{ S_{\a},T_{\b}\}&=\e_{\a\b\g}T_{\g},\\
\{ T_{\a}, T_{\b}\}&=0.
\end{align*}
The question is to know how the roots of the characteristic equation attached to the Clebsch system behave with respect to this bracket. We prove that they are in involution. Thus we have the following stronger statement.
\begin{teo}
The Clebsch system is a KCI system.
\end{teo}
\textit{Proof}. We want  to prove that $\{x_1,x_2\}=0 $. Since $x_1=v$, $x_2=-v-F/E$, this is equivalent to prove that
\begin{equation*}
F\{v,E\}=E\{v,F\}. \
\end{equation*}
We will prove the stronger conditions $\{v,E\}=0$ and $\{v,F\}=0$.
Let us show at first that  $\{v,E\}=0$. We have:
\begin{equation*}
\label{comve}
\{v,E\}=\sum\limits_{\a=1}^3c_{\a}v_{\a}\{v,T_{\a}\}=\sum\limits_{\a=1}^3c_{\a}v_{\a}(\frac{\partial
v}{\partial S_{\g}}T_{\b}-\frac{\partial v}{\partial
S_{\b}}T_{\g}).
\end{equation*}
To proceed we need to know  $\dfrac{\partial v}{\partial S_{\g}}$. This derivative is furnished by the Implicit Function Theorem :
\begin{equation*}
\dfrac{\partial v}{\partial S_{\a}}=\l c_{\a} v_{\a} \qquad \l=\dfrac{-2v_1v_2v_3}{E(x_1-x_2)} \
\end{equation*}
Hence the Poisson bracket $\{v,E\}$ is a rational function of the
variables ${S}_{\a}$,${T}_{\a}$, ${v}_{\a}$, $v$ ( as explained at
the beginning of this section). This function, however, vanishes
on account of the relations $v_{\a}^2=v+j_{\a}$,
$c^2_{\a}=j_{\b}-j_{\g}$ and formulas (\ref{reconstr}). Thus
$\{v,E\}=0$. In the same manner one proves that $\{v,F\}=0$.
 Therefore the characteristic equation attached to the Clebsch system satisfies the third Kowalewski's condition. Since it verifies also the first two conditions according to the previous Theorem, it remains proved that the Clebsch system is a KCI system.
\begin{flushright}
$\square$
\end{flushright}

The second aim of this section is to explain where the characteristic equation attached to the Clebsch system comes from.
 This explanation is splitted in a series of eight short remarks, presented here without proofs. The proofs are
  omitted since the following discussion has merely an illustrative purpose, and since these remarks will never be
  used later on in the paper. We stress that in this subsection we are working within the framework of the Hamiltonian
   formulation of the Clebsch system.

\bigskip
\noindent
\textbf{Remark 1} . The simplest way to arrive to the characteristic equation attached to the Clebsch system is to exploit a third possible form of the Kowalewski's conditions. It concerns the coefficients $f=\dfrac{F}{E}$, $g=\dfrac{G}{E}$ of the characteristic equation reduced to  monic form. We claim that they satisfy the conditions
\begin{align*}\label{preKow}
\begin{split}
\textit{K1''}: \qquad  f'&=\dot{g},\\
\textit{K2''}: \qquad g'&=f\dot{g}-g\dot{f}.\\
\textit{K3''}: \qquad \{f,g\}&=0.
\end{split}
\end{align*}
These conditions will be referred to as the third form of the Kowalewski's conditions, and called  the conditions $\textit{K1''}$ , $\textit{K2''}$ , $\textit{K3''}$  respectively.

\bigskip
\noindent
\textbf{Remark 2} . This remark explains how to construct solutions of  conditions $\textit{K1''}$ and $\textit{K3''}$ by taking the derivatives of the Hamiltonian functions $H$ and $K$ along a suitably chosen vector field $Z$. We claim that if $Z$ satisfies the conditions
\begin{equation*}
Z^2(H)=0, \  Z^2(K)=0
\end{equation*}
\begin{equation*}
Z \{A , B \}- \{ZA , B \}- \{A , ZB \}=0.
\end{equation*}
$ \forall A,B \in C^{\infty}(M)$, then the functions
\begin{equation*}
f=Z(H), \ g=Z(K)
\end{equation*}
satisfy the first and the third conditions presented above. The last condition means that the vector field $Z$ is a symmetry of the Poisson bracket of the Clebsch system.

\bigskip
\noindent
\textbf{Remark 3} . This remark specifies the action of the vector field $Z$ on the Casimir functions.
 It is known that a symmetry of a Poisson bracket transforms Casimir functions into Casimir functions. Therefore the functions $Z(C_1)$ and $Z(C_2)$ must be still Casimir functions. The simplest possibility is that they be constants. Accordingly we impose the supplementary conditions
 \begin{equation*}
Z(C_1)=a, \  Z(C_2)=b,
\end{equation*}
on $Z$, $a$, $b$ being arbitrary constants.

\bigskip
\noindent
\textbf{Remark 4} . This remark explains how to find the vector fields $Z$ which verify the four conditions
\begin{equation*}
Z^2(H)=0, \  Z^2(K)=0, \ Z(C_1)=a, \  Z(C_2)=b,
\end{equation*}
These conditions specify the action of the vector field $Z$ on the integrals of the foliation $\mathcal{F}$
associated to the Clebsch system. Since we also know that $Z$ is a symmetry of the Poisson bracket, we can infer from them the action of $Z$ on the vector fields $X$ and $Y$ as well. Therefore, through these conditions we can completely control the action of $Z$ on  the foliation attached to the Clebsch system. Our main remark is that the  vector field
\begin{equation*}
Z=\frac{b}{2(\sum\limits_{\a}^3
c_{\a}v_{\a}T_{\a})}(\sum\limits_{\a=1}^3
c_{\a}v_{\b}v_{\g}\frac{\partial }{\partial S_{\a}}+
\sum\limits_{\a=1}^3 c_{\a}v_{\a}\frac{\partial }{\partial
T_{\a}}),
\end{equation*}
satisfies the above conditions if $v$ satisfies the equation
\begin{equation*} \label{Vinc}
 b\sum c_{\alpha}(v_{\alpha} v_{\gamma}T_{\alpha}+v_{\alpha}S_{\alpha})=2a\sum c_{\alpha}v_{\a}T_{\a}
\end{equation*}

\bigskip
\noindent
\textbf{Remark 5} . This remark adds the information that all the vector fields $Z$ just found are symmetries of the Poisson bracket of the Clebsch system.  Therefore we know, at this point,  a 2-parameter family of solutions $f=\dfrac{F}{E}$, $g=\dfrac{G}{E}$ of  conditions $\textit{K1''}$, $\textit{K3''}$ listed before.

\bigskip
\noindent
\textbf{Remark 6} . It remains to study the second Kowalewski's condition $\textit{K2''}$. It fixes the value of the parameter $a$. We claim that the  vector field $Z$ satisfies the second Kowalewski's condition if and only if  $a=0$.

\bigskip
\noindent
\textbf{Remark 7} . Without  loss  of generality one can put $b=1$. We thus arrive at the following conclusion:  the functions
\begin{align*}
f&=\frac{\sum\limits_{\a=1}^3c_{\a}(v_{\b}v_{\g}S_{\a}+(j_{\b}+j_{\g})v_{\a}T_{\a})}{\sum\limits_{\a=1}^3c_{\a}v_{\a}T_{\a}},\\
g&=\frac{\sum\limits_{\a=1}^3c_{\a}(v_{\b}v_{\g}j_{\a}S_{\a}+j_{\b}j_{\g}v_{\a}T_{\a})}{\sum\limits_{\a=1}^3c_{\a}v_{\a}T_{\a}}
\end{align*}
satisfy all the Kowalewski's conditions if the function $v$ is a solution of the equation
\begin{equation*} \label{Vinc}
\sum c_{\alpha}(v_{\alpha} v_{\gamma}T_{\alpha}+v_{\alpha}S_{\alpha}) =0 .
\end{equation*}

\bigskip
\noindent
\textbf{Remark 8}. Finally one passes to the form of the characteristic equation presented in the Introduction  by multiplying the coeffients $f$ and $g$ of the monic characteristic equation by their common denominator $E=\sum\limits_{\a=1}^3c_{\a}v_{\a}T_{\a}$.

\section{The Abel and the Jacobi forms of the Clebsch equations}

We are now ready to prove the  theorems  stated in the Introduction, relative to the Abel's
form and to the Jacobi's form of the equations of motion of the Clebsch system.

As a first step, we write the integrals $(I_1, I_2, I_3, I_4)$  of the foliation $\mathcal{F}$ associated
with  the Clebsch system in the intermediate coordinates $x_1$, $x_2$, $e$, $l_1$, $l_2$, $l_3$.  A direct calculation
 gives:
\bigskip
\begin{small}
\begin{align*}
\begin{split}
I_1&=-el_2,\\
I_2&=-(e^2(j_1+j_2+j_3+x_1+2x_2) +
2el_3+\bigl(\frac{l_1-l_2}{x_1-x_2}\bigr)^2),\\
I_3&=(e^2(-j_1j_2-j_2j_3-j_3j_1+x_2^2+2x_1x_2)+e(-2\sqrt{(x_1+j_1)(x_1+j_2)(x_1+j_3)}\frac{l_1-l_2}{x_1-x_2}+\\
&+2(x_1+x_2)l_3)+2(l_1-l_2)\frac{(l_1x_2-l_2x_1)}{(x_1-x_2)^2}),\\
I_4&=(-e^2(j_1j_2j_3+x_1x_2^2)+ex_2(2\sqrt{(x_1+j_1)(x_1+j_2)(x_1+j_3)}\frac{l_1-l_2}{x_1-x_2}-2l_3x_1)+\\
&-\frac{(l_1x_2-l_2x_1)^2}{(x_1-x_2)^2}).
\end{split}
\end{align*}
\end{small}

As a second step we invert these formulas. To simplify the inversion it is useful to notice that
the equations for $\dot{x}_1$ and $\dot{x}_2$  are independent of the coordinate $l_3$.
Therefore we are interested in solving the above equations only with respect to $e$, $l_1$,
$l_2$. Since $I_1$ does not contain $l_3$, the first problem is to eliminate $l_3$ from the other equations . This result is easily achieved by introducing the functions:
\begin{align*}
E_1&= I_2 x_1^2+I_3x_1+I_4 ,  \nonumber\\
E_2&= I_2 x_2^2+I_3x_2+I_4 .  \nonumber\
\end{align*}
Adding the equation for $I_1$ and substituting the above expressions for the integrals $I_1$, $I_2$, $I_3$, one obtains a system of three equations for three unknowns
\bigskip
\begin{small}
\begin{align*}
\begin{split}
I_1&=-el_2,\\
E_1&=-(2C_1
\sqrt{(x_1+j_1)(x_1+j_2)(x_1+j_3)}+(e\sqrt{(x_1+j_1)(x_1+j_2)(x_1+j_3)}+l_1)^2),\\
E_2&=-(e^2(x_2+j_1)(x_2+j_2)(x_2+j_3)+l_2^2),
\end{split}
\end{align*}
\end{small}
\bigskip
These equations give the  desired answer for $e$, $l_1$, $l_2$.
\bigskip

The final step is to insert the expressions for $e$, $l_1$, $l_2$ into the equations of motion for $x_1$ and $x_2$  already computed in the previous section.
\begin{lem}
In the final coordinates $x_1$, $x_2$, $I_1$, $I_2$, $I_3$, $I_4$ the equations of motion of the Clebsch  system take the form:
\bigskip
\begin{footnotesize}
\begin{align*}
\begin{split}
(x_1-x_2)\dot{x}_1&=-{4i}\sqrt{(x_1+j_1)(x_1+j_2)(x_1+j_3)}
\sqrt{x_1^2I_2+x_1I_3+I_4+2 I_1\sqrt{x_1+j_1}\sqrt{x_1+j_2}\sqrt{x_1+j_3}},\\
\bigskip
(x_2-x_1)\dot{x}_2&={\sqrt{2}\sqrt{x_2+j_1}\sqrt{x_2+j_2}\sqrt{x_2+j_3}} \times \\
&\times\Bigl(-
\sqrt{-(x_2^2I_2+x_2I_3+I_4)+\sqrt{(x_2^2I_2+x_2I_3+I_4)^2-4I_1^2(x_2+j_1)(x_2+j_2)(x_2+j_3)}}+\\
&+4I_1^2\frac{(x_2+j_1)(x_2+j_2)(x_2+j_3)}{(-(x_2^2I_2+x_2I_3+I_4)+\sqrt{(x_2^2I_2+x_2I_3+I_4)^2-
4I_1^2(x_2+j_1)(x_2+j_2)(x_2+j_3)})^{\frac{3}{2}}}\Bigr).
\end{split}
\end{align*}
\end{footnotesize}
\bigskip
\end{lem}
\textit{Proof}. These formulas are the outcome of the elimination of $e$, $l_1$, $l_2$ among the the three equations for $I_1$, $E_1$, $E_2$  and the two equations for $\dot{x}_1$ and $\dot{x}_2$ found in the previous section.
\begin{flushright}
$\square$
\end{flushright}

By this Lemma we reach the central point of our study : the identification of the pair of functions $\psi_1(x_1,I)$ and $\psi_2(x_2,I)$ whose existence was implied by the Kowalewski's property. The deduction of these functions essentially completes the study of the equations of motion of the Clebsch system. All the other inferences are simple consequences of this result. We collect them in the form of  corollary of what has been said so far.Let us recall the $I_1=C_1$, $I_2= C_2$, $I_3 = H$,$I_4 = K$.

\begin{nas}
The functions  $\psi_1(x_1,I)$ and $\psi_2(x_2,I)$ of the Clebsch system are the restriction of the rational functions $R_{1}( x_{1}, y_{1} )$ and $R_{2}( x_{2}, y_{2} )$ presented in Sec. 1, on the algebraic curves $\g_{1}$ and $\g_{2}$ shown there. Furthermore, the canonical momenta conjugated to the roots of the characteristic equation attached to the Lagrangean foliation $\mathcal{F}$ of the Clebsch system are
\begin{small}
\begin{align*}
4p_1^2(x_1+j_1)(x_1+j_2)(x_1+j_3)+(x_1^2C_2+x_1H+K)+2\sqrt{x_1+j_1}\sqrt{x_1+j_2}\sqrt{x_1+j_3}C_1&=0,\\
p_2^2(x_2+j_1)(x_2+j_2)(x_2+j_3)+\frac{C_1^2}{p_2^2}+
(x_2^2C_2+x_2H+K)&=0.
\end{align*}
\end{small}
\end{nas}
\textit{Proof}. The first claim is proved by a simple checking. The second  claim is the outcome of the algorithm for the computation of canonical momenta presented in Theorem 2.2.
\begin{flushright}
$\square$
\end{flushright}

With the computation of the rational functions $R_{1}( x_{1}, y_{1} )$ and $R_{2}( x_{2}, y_{2} )$, of the algebraic curves $\g_{1}$ and $\g_{2}$, and of the conjugate momenta $p_1$ and $p_2$ the Theorems stated in Sec.1 are completely proved. To these results we simply add two minor remarks : the form of the Hamiltonian functions in the canonical coordinates
\begin{footnotesize}
\begin{align*}
\begin{split}
H&=\frac{4p_1^2(x_1+j_1)(x_1+j_2)(x_1+j_3)-(p_2\sqrt{(x_2+j_1)}\sqrt{(x_2+j_2)}\sqrt{(x_2+j_3)}-\dfrac{C_1}{p_2})^2}{x_2-x_1}-(x_2+x_1)C_2\\
\bigskip
K&=\frac{4x_2p_1^2(x_1+j_1)(x_1+j_2)(x_1+j_3)-x_1(p_2\sqrt{(x_2+j_1)}\sqrt{(x_2+j_2)}\sqrt{(x_2+j_3)}-\dfrac{C_1}{p_2})^2}{x_1-x_2}+
x_2x_1C_2+\\
&+2\frac{x_2\sqrt{(x_1+j_1)}\sqrt{(x_1+j_2)}\sqrt{(x_1+j_3)}-x_1\sqrt{(x_2+j_1)}\sqrt{(x_2+j_2)}\sqrt{(x_2+j_3)}}
{x_1-x_2}C_1.
\end{split}
\end{align*}
\end{footnotesize}
and that of the original mechanical variables ${S}_{\a}$, ${T}_{\a}$ . In this second case, the variables are given by the equations
\begin{small}
\begin{align*}
\begin{split}
S_{\a}&=\frac{c_{\a}}{c_1c_2c_3}(
\sqrt{x_1+j_{\a}}\frac{(j_{\b}+j_{\g}+x_1+x_2)}{x_1-x_2}l_1-
\sqrt{x_1+j_{\a}}\frac{(j_{\b}+j_{\g}+2x_1)}{x_1-x_2}l_2\\
&-\sqrt{x_1+j_{\b}}\sqrt{x_1+j_{\g}}l_{3}),\\
\bigskip
T_{\a}&=\frac{c_{\a}}{c_1c_2c_3}(
\frac{\sqrt{x_1+j_{\b}}\sqrt{x_1+j_{\g}}}{x_1-x_2}(l_{1}-l_2)-
\sqrt{x_1+j_{\a}}l_3-\sqrt{x_1+j_{\a}}(x_2+j_{a})e).
\end{split}
\end{align*}
\end{small}
after the replacement of the intermediate coordinates  $l_1$, $l_2$, $l_3$ , $e$ by their expressions in terms of the canonical coordinates:
\begin{align*}
\begin{split}
e=& p_2,\\
l_1=&
\sqrt{x_1+j_1}\sqrt{x_1+j_2}\sqrt{x_1+j_3}(2p_1-p_2), \\
l_2=&
-\frac{C_1}{p_2}\\
l_3=&-
\frac{1}{2p_2}(C_2+p_2^2(j_1+j_2+j_3+x_1+2x_2)+\frac{(l_1-l_2)^2}{(x_1-x_2)^2}).
\end{split}
\end{align*}

Since the Jacobi's  theorem allows  to evaluate, in principle, the dependence of the canonical coordinates  on the time, these reconstruction formulas allow one to present a sufficiently accurate picture of the dynamics of the Clebsch system in the physical space. At this point, one may consider  the study of  the dynamics of the Clebsch system as reasonably completed.

\section{Concluding Remarks}

The first systematic study of the Clebsch system has been performed, more than one century ago, by Kotter in a monumental paper published in 1892 \cite{Kotter}. In that paper he managed to find a complicated change of variables that brings the quadratic integrals of motion of the Clebsh system in a form (called the " canonical quadrics"  of Kotter  by Perelomov \cite{Perelom} ) which suggests a certain  similarity of the behaviour of the Clebsch system with the behaviour of a particle on the unit sphere $\mathbb{S}^{2}$  under the action of a quadratic potential. The idea of the similarity between the Clebsch system and the Neumann system  \cite{Neu}  was rather natural, since this similarity becomes an  homomorphism in the particular case of the leaves of the foliation $\mathcal{F}$ corresponding to the special value $C_1=0$ and $C_2=1$ of the Casimir functions. The symplectic manifold corresponding to these values of the Casimir functions have, indeed, the topology of the cotangent bundle of the unit sphere  $\mathbb{S}^{2}$. It is difficult nowadays to read and to understand the paper by Kotter. A recent attempt has been done, from the viewpoint of algebraic geometry,  by Enolski and Fedorov \cite{EnolFed} , from who we have learnt that the main tool used by Kotter to find the solutions of the Clebsch system is an algebraic curve of genus two. This immediately points out a striking difference between the present approach and that of Kotter. In our approach we need a couple of algebraic curves of genus three to write the equations of motion in the Abel form, while Kotter makes use of a single curve of genus two. We are unable to explain this discrepancy. It is certainly possible that the same system may be brought into the Abel form in several distinct ways, but one would like to control the relationship that should exists among these different ways. We think that it could be interesting to analyse the approach of Kotter from the viewpoint of the theory of dynamical systems which are integrable in the sense of Lie, introduced in the present paper. Kotter, as Kowalewski few years before, never consider explicitly the symmetry $Y$ associated with the Clebsch sytem $X$. In this way he forbids himself to exploit an important piece of information about the Clebsch system. Our guess is that the analysis of the derivatives of the Kotter transformation along the vector fields $X$ and $Y$ should give the hint to understand the relationship among the approach of Kotter and that proposed in this paper. This guess, however, at the moment is simply an open question.

Many other open questions are raised by the comparison with other, more recent, attempts of describing the properties of the Clebsch system \cite{AdlerVanMoerbeke}. After the new concept of Lax representation has become central in the theory of integrable systems, many efforts have been spent to identify the class of integrable systems which admit a Lax representation, and whose flow can be linearized on the Jacobian ( or Prym ) variety of the spectral curve provided by the Lax matrix \cite{Babelon}. The name of  ACI systems, or algebraic completely integrable systems, have been coined to denote this class of systems \cite{AdlerVanMoerbekeVanhaecke}. In our opinion the ACI systems are closely related ( possibly coincide) with the integrable systems, in the sense of Lie, which are related to foliations $\mathcal{F}$ possessing a characteristic equation which enjoys the Kowalewski's property. The study of the Clebsch system is quite inspiring  from this point of view. The Clebsch system has indeed two  different Lax representations, and therefore two spectral curves. Both  curves appear in the Abel form of the equations of motion elaborated in this paper.They are the algebraic curves $\g_{1}$ and $\g_{2}$  described in the Introduction ( up to a suitable rescaling of the Lax matrix). This notice clarifies the remark made before on the relation between the integrability in the sense of Lax and in the sense of Lie, and set a question: Why in the standard theory of ACI systems the attention is adressed to a single spectral curve at a time? In the example of the Clebsch system, as treated in this paper, two spectral curves appear indeed simultaneously. We believe that this example should provide the imput for a critical rethinking of the foundations of the theory of ACI systems.

Once the Hamiltonian structure of the equations of motion is taken into account, the theory of ACI systems merges
with the theory of separable Hamiltonian systems. This is the point of view emphasized by  Sklyanin \cite{Sklyanin}.
The spectral curve of the Lax matrix  plays, in this context, the role of  " separation equation " . Normally, a separable
 Hamiltonian system has many separation equations ( as many as the degrees of freedom), and only in special ( degenerate)
  cases these equations coincide, so that a single separation equation eventually survives. This remark reinforces the
   question set before: Why to consider a single spectral curve at a time? The example of the Clebsch system shows that
   the standard assumptions of the Lax approach to separation of variables probably need to be weakened. Another feature
    of the Lax approach to separation of variables deserves a few comments. The Lax matrix provides the separation
    equation, but the separation coordinates should be produced otherwise. For this purpose, the analysis of the Lax
    matrix must be complemented with the study of  the singularities of the " Baker-Akhiezer function" . We don't like to
     discuss this part of the theory here, but we notice that its purpose  is to provide two meromorphic functions
      $A( u ) $ and $ B( u) $ of a complex variable $u$ (the spectral parameter of the Lax matrix ), which verify two
       conditions : 1) the separation coordinates can be found among the zeroes of $B( u ) $ ; 2 ) the values of $A( u ) $
        on these zeroes are the conjugate momenta. The analogy of the role of the meromorphic function $B( u ) $ with the
         role of the  characteristic equation attached to the foliation $\mathcal{F}$ is obvious, but to work out this
         analogy in detail is a demanding task. The most interesting open problem  is to implement the Kowalewski's
         conditions, presented in Sec. 2, onto the meromorphic function  $B( u ) $.  Another interesting problem is
         to interpret as well the second meromorphic function  $A( u ) $ from the viewpoint of Theorem 2.2.How is it
          possible to define this function when two spectral curves are simultaneously present? More generally: What is the relation between the Hamiltonian systems which are separable in the sense of Sklyanin and the KCI systems ? We leave this question as an open problem.

The bihamiltonian approach to integrable systems often provides an alternative way for the study of  separation of variables. When the Poisson pencil is of the type analysed by Gel'fand- Zakharevich \cite{GelZakh}, one can follow the procedure outlined in \cite{FalquiPedroni}. The bihamiltonian  scheme  concerns a more restricted class of dynamical systems than that considered in this paper, but it may be of use in the search of the characteristic equation when the dynamical system satisfies the more restrictive conditions. The Clebsch system is a good test for this theory as well. Indeed the Clebsch system is bihamiltonian. We have verified that the bihamiltonian recipe suggested in \cite{FalquiPedroni} is a special case of the Kowalewski's conditions used in this paper.

The last approach to the Clebsch system which is worth to be mentioned is the " partial separation of variables" proposed in \cite{MarSok}. At first sight,  this scheme is different from the present one, even if one may recognizes several points of contact. The main difference is that the authors of that paper use three Abelian differentials on the curve $\g_2  $, while we use two Abelian differentials on two curves. We admit, however, that we lacked the time needed  for a careful comparison between the two approaches, and therefore also this question is left as an open problem.

We hope that this cursory glance at the different ways of looking at the Clebsch system  will serve to increase the interest for a comparative study of the many partial viewpoints  encountered in the theory of integrable systems, both in its classical and modern form.

\end{document}